\documentclass[letterpaper, 10pt, conference]{ieeeconf}

\usepackage{cite}
\usepackage{amsmath}
\usepackage{amssymb}
\usepackage{hyperref}
\usepackage{caption}
\usepackage{subcaption}
\usepackage[short,c3]{optidef}

\newcommand*{\T}{\mathsf{T}}
\newcommand*{\R}{\mathbb{R}}
\newcommand{\bmat}[1]{\begin{bmatrix}#1\end{bmatrix}}
\newcommand{\uVec}{\mathbf{u}_{0:N-1}}
\newcommand{\xVec}{\mathbf{x}_{0:N}}
\newcommand{\J}{J(\xVec, \uVec)}

\title{\LARGE \bf
On the Application of Model Predictive Control to a Weighted Coverage Path Planning Problem
}
\author{Kilian Schweppe, Ludmila Moshagen, Georg Schildbach}

\begin{document}

\maketitle
\thispagestyle{empty}
\pagestyle{empty}

\begin{abstract}
This paper considers the application of Model Predictive Control (MPC) to a weighted coverage path planning (WCPP) problem.
The problem appears in a wide range of practical applications, including search and rescue (SAR) missions.
The basic setup is that one (or multiple) agents can move around a given search space and collect rewards from a given spatial distribution.
Unlike an artificial potential field, each reward can only be collected once.
In contrast to a Traveling Salesman Problem (TSP), the agent moves in a continuous space.
Moreover, he is not obliged to cover all locations and/or may return to previously visited locations.
The WCPP problem is tackled by a new Model Predictive Control (MPC) formulation with so-called Coverage Constraints (CCs).
It is shown that the solution becomes more effective if the solver is initialized with a TSP-based heuristic.
With and without this initialization, the proposed MPC approach clearly outperforms a naive MPC formulation, as demonstrated in a small simulation study.
\end{abstract}

\section{Introduction}

State-of-the-art methodologies in robotic path planning typically seek to determine an optimal trajectory, maximizing a reward function given the system's dynamic capabilities. The reward function is often a weighted sum of terms for particular purposes, e.g., proximity-based guidance by Artificial Potential Fields (APF) or enforcing safety boundaries via barrier functions. Model Predictive Control (MPC) fully encompasses this framework by formulating the problem as a numerical optimization program. Over the past decade, MPC has become a standard approach for many path planning applications, owing to its inherent capacity to effectively integrate general objective functions, complex system dynamics, and constraints on the system's inputs and states.

Coverage Path Planning (CPP) is characterized by the mandate for the system to traverse, or cover, the entire area of the designated state space, instead of moving to a target area or point. Instances of this problem are prevalent in diverse robotic domains, including autonomous agricultural systems, automated cleaning devices, and aerial or underwater drones employed for inspection or surveillance. Motion planning approaches are typically based on pre-defined patterns or simple policies, including, but not limited to, boustrophedon (snake-like) patterns or random-walk strategies involving boundary reflections \cite{hasan2014path}. The core principle underpinning these approaches is the minimization of spatial and temporal overlaps in the system's trajectory.

Weighted Coverage Path Planning (WCPP) is an extension of CPP. The primary difference is the reward is not uniformly distributed, as for the CPP problem. Instead, it is weighted by an explicit coverage priority. Example applications include 
\begin{itemize}
  \item[(i)] a mobile irrigation system, which may aim to prioritize areas based on their moisture level, preferring areas with particularly dry soil;
  \item[(ii)] a cleaning robot, whose goal may be to collect as much dirt as possible, hence using a coverage priority based on the actual distribution of dirt;
  \item[(iii)] an aerial or underwater inspection or surveillance system, whose priority distribution is related to the probability of detecting a fault or finding a target in a given area or on a given map.
\end{itemize}
CPP problems usually assume a static environment, such as the cleaning of a room. WCPP problems, on the other hand, often appear in dynamic settings, such as the surveillance for moving targets. Clearly, accounting for dynamically changing environment adds a extra level of complexity, and it is also possible to consider WCPP for a static, yet not uniform coverage priority.

This paper focusses on the static WCPP. It is motivated by search and rescue (SAR) missions, e.g., in the Wadden Sea, using an Unmanned Aerial Vehicle (UAV). The goal is to locate a lost person as quickly as possible. Thus the coverage priority is given by a belief distribution of the person's current location, as a function over the search area. This so-called \emph{probability map} may be based on information about the last known location of the person, a behavioral model, the environmental conditions, and the elapsed time \cite{lin2010bayesian}. More details on the generation of the probability map and a comparison of search methods can be found in a recent publication \cite{MoshagenEtAl:2025}.

This paper proposes an MPC formulation for the static WCPP. From the perspective of MPC, the priority function, or probability map, can be considered as a reward function, or a negative cost function. The major difference to conventional MPC is that each reward can only be collected once, because upon entering a certain position, the related priority drops to zero. In the case of the SAR mission, for example, the UAV is supposed to plan a trajectory to collect as much of the probability mass as possible, instead of hovering above the point with the highest probability value.

\subsection{Literature Review}

Many algorithms have been proposed for CPP in different environments with and without obstacles \cite{galceran2013survey,tan2024comprehensive}. Algorithms can be divided into randomized and deterministic, and they are based on techniques such as AFPs, greedy search, graph search algorithms, and bio-inspired approaches, like genetic algorithms or ant colony optimization.

A key point for some applications is that the resulting paths should be smooth \cite{tan2024comprehensive}. In fact, some robots are not able to turn on the spot. Sharp turns may also slow the robot down or lead to premature wear of the robot's components. For example, in \cite{tripicchio2023smooth}, a search algorithm based on ant colony optimization with the Lin-Kernighan heuristic, is combined with a path smoothing using Fourier series. The resulting trajectory is then fed to the UAV, which is operated by a controller based on MPC. 

MPC itself has been successfully employed for collision-free path planning for autonomous road vehicles \cite{RaseEtAl2017} as well as UAVs \cite{Woods2017Novel}. It has been combined with with AFPs \cite{StoicanEtAl2022,ZhuEtAl2024}, and it has also been employed previously for CPP with obstacles using a mixed-integer linear programming (MILP) formulation \cite{ibrahim2019hierarchical}. To this end, the area is covered by a uniform grid, where each cell defines a single way point. The way points are subsequently represented as discrete decision variables within the optimization problem, where the objective is to cover the maximum number of (equally weighted) way points.

Related to CPP and WCPP is the Traveling Salesman Problem (TSP) and its variants, most notably the Orienteering Problem (OP) \cite{golden1987orienteering}. In contrast to the TSP, where the salesman has to visit all vertices of the graph and the goal is to minimize the traveled distance, the OP is concerned with maximizing the total reward collected within a limited time frame, without necessarily visiting all the vertices \cite{orienteering}. Both problems are known to be NP-hard~\cite{lin2009uav}. Yet it is desirable, for many applications, to extend them by considering a spacial map instead of a graph, and including the dynamics of an agent in the problem formulation \cite{Ross_2019b}. A mathematical framework to tackle this problem using techniques from optimal control has been recently proposed \cite{Ross_2019b, Ross_2020a}. In addition to the dynamics of an agent, this formulation also accounts for the movement of a sensor, in this case, a camera. However, the approach relies on non-smooth calculus and considers only discrete regions of interest.

\subsection{Contributions}

This paper proposes a new MPC approach for the WCPP problem, using \emph{coverage constraints} (CCs).
The proposed MPC formulation works with any agent moving governed by a nonlinear dynamic model and moving in a continuous space with obstacles.

The reward function is arbitrary, but in contrast to an AFP, each reward can \emph{only be collected once}. This is enforced by the use of quadratic constraints. In contrast to the TSP, the agent is not obliged (and may in fact be far from able to) cover all locations within the given prediction horizon.

Furthermore, the paper shows that the solution becomes more effective if the MPC solver is initialized with a \emph{TSP-based heuristic}. The heuristic is based on a set of key points, which are derived from a Gaussian mixture model that is used to approximate the reward function.

\section{Methods}\label{methods}


Consider a reward function $r: \mathbb{R}^n \to \mathbb{R}$, which maps a state $x \in \mathbb{R}^n$ to its reward $r(x)$.
It is assumed that $r$ is twice continously differentiable, in order to enable gradient-based optimization.
The goal is to find a trajectory $\bar{h}: [t_0,t_1] \to \mathbb{R}^n$ with $\bar{h}(t_0) = x_{0} \in \mathbb{R}^n$ satisfying the dynamic constraints $\dot{x} = \bar{f}(x,u)$ of an agent. The objective is to maximize the reward along the path by choosing the control input $u \in \mathbb{R}^m$ between time $t_{0}$ and $t_{1}$. 
Specifically, we seek to maximize the integral
\begin{equation}\label{eq:integral}
\int_{t_{0}}^{t_{1}} r\left(\bar{h}(t)\right) \mathrm{d}t
\end{equation}
As shown in Section~\ref{impl1}, the objective (\ref{eq:integral}) can be easily incorporated into the objective function of the MPC.
However, as the problem is highly non-linear in the input signal $u(t)$, we try to improve the solution by providing a good initial guess to the NLP solver. 
The problem is therefore solved in a hierachical fashion, consisting of multiple steps:
\begin{enumerate}
        \item Find a set of key points using Gaussian mixture models as described in Section~\ref{impl2};
        \item Find an optimal path going through all key points using a travelling salesman formulation as described in Section~\ref{impl2};
        \item Solve the receding horizon optimal control problem as explained in Section~\ref{impl1}, using the solution obtained in the previous step as an initial guess.
\end{enumerate}
Note that in the case of a discrete probability grid, the continuous reward function can be constructed based on a B-spline interpolation, for example.

\subsection{Finding a trajectory with MPC}\label{impl1}

We consider a discrete-time optimal control problem of finding the optimal sequence of control inputs $u_{0},u_{1},…,u_{N-1} \in \mathbb{R}^m$, driving the states $x_{0},x_{1},…,x_{N} \in \mathbb{R}^n$ of a dynamical system $x_{k+1} = f(x_{k},u_{k})$ for $k=0,1,…,N-1$, where $N$ is the horizon length.

Denote by $\uVec := \{u_0,u_1,…,u_{N-1}\}$ the sequence of control inputs and by $\xVec := \{x_0,x_1,…,x_{N}\}$ the corresponding sequence of states. Let $r(x_k)$ be the reward at state $x_k$.
Then the (non-linear) optimal control problem is given by
\begin{mini!}
{\xVec,\uVec}{\J\,, \label{prob:obj}}
{\label{prob}}{}
\addConstraint{x_{k+1}}{= f(x_k,u_k) \quad \label{prob:constr:1}}{\forall\;k=0,…,N-1}
\addConstraint{x_k \in \mathbb{X}}{\label{prob:constr:3}}{\forall\;k=0,…,N}
\addConstraint{u_k \in \mathbb{U}}{\label{prob:constr:4}}{\forall\;k=0,…,N-1}
\addConstraint{x_0 = x(0)\;.}{\label{prob:constr:5}}
\end{mini!}
The objective function $\J$ contains the control input costs and the state rewards:
\begin{equation}
    \J = c_1 \sum_{k=0}^{N-1} u_k^{\T} R u_k - c_2 \sum_{k=0}^{N} r(x_k)\;.
\end{equation}
For the control input costs we choose a quadratic formulation, defined by a positive definite cost matrix $R \in \R^{m \times m}$.
The second term in the objective function maximizes the reward along the path, approximating the integral in (\ref{eq:integral}).
Both terms are weighted by the (positive) parameters $c_1$ and $c_2$, respectively.

The constraints (\ref{prob:constr:3}) and (\ref{prob:constr:4}) keep the states and inputs in the set of admissible states (the region of the map) \(\mathbb{X}\) and in the set of admissible inputs (accelerations) \(\mathbb{U}\), respectively.
The constraint (\ref{prob:constr:5}) sets the initial state of the agent.

To prevent the agent from just collecting the same reward multiple times, we introduce additional \emph{coverage constraints}
\begin{equation}\label{prob:constr:2}
    \rho(x_{i}, x_{j}) \geq V - \epsilon_i, \quad i=1,…,N,~j=0,…,i-1
\end{equation}
for some distance metric $\rho(x_i,x_j)$.
These constraints force the agent to move by ensuring that each state $x_i$ along the trajectory is at least distance $V$ away from each previous state $x_j$ for $j < i$.
The parameter \(V\) is related to the \textit{visibility} of the agent, i.e., it determines how far the agent needs to travel in order to collect a new reward.
Note that the constraint is implemented as a soft constraint with the slack variables $\epsilon_i \in \R$ in order to allow the agent to slightly violate a CC if this leads to a better trajectory.
The slack variables appear as an additional term in the modified objective function
\begin{multline}
    \J =\\ c_1 \sum_{k=0}^{N-1} u_k^{\T} R u_k - c_2 \sum_{k=0}^{N} r(x_k) + c_3 \sum_{k=0}^{N} \epsilon_k\;.
\end{multline}
Note that the dynamics (\ref{prob:constr:1}) in the MPC formulation are generally arbitrary. For the implementation, simple linear dynamics \(f(x_{k},u_{k}) = A x_{k} + B u_{k}\) are assumed, in the form of a double integrator:
\[A = \bmat{1 & 0 & \Delta t & 0 \\ 0 & 1 & 0 & \Delta t \\ 0 & 0 & 1 & 0 \\ 0 & 0 & 0 & 1},~B = \bmat{0 & 0 \\ 0 & 0 \\ \Delta t & 0 \\ 0 & \Delta t}\]\;.
Here, the states are the positions and velocities, and the inputs are the accelerations.
The function $\rho(x_i,y_i)$ is defined as the distance between the positions of two states, i.e.,
\begin{equation}
    \rho(x_i,x_j) = \sqrt{(x_{i,1} - x_{j,1})^2 + (x_{i,2} - x_{j,2})^2}\;.
\end{equation}
This leads to the quadratic CCs
\begin{multline}
     (x_{i,1}-x_{j,1})^2 + (x_{i,2} - x_{j,2})^2 \geq V^2 - \epsilon_i \\ \forall\; i=1,…,N\;\text{and}\;j=0,…,i-1\;.
\end{multline}

\begin{figure}
\includegraphics[width=\linewidth]{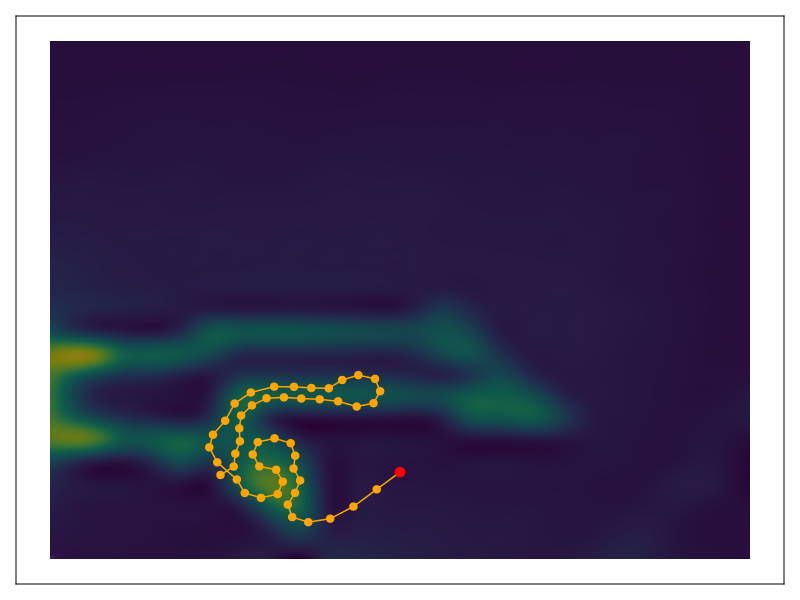}
\caption{Solution of the MPC for the same scenario as in Figure~\ref{fig:results:1}, but without an initial guess.}
\label{fig:noguess}
\end{figure}

\subsection{Finding an initial solution with a TSP based heuristic}\label{impl2}

Depending on the particular problem instance, the MPC with CCs may itself already yield satisfactory results for the WCPP problem.
However, it can be observed that in some cases the solution gets stuck in a local minimum, leading to a sub-optimal behavior.
An example of this is shown in Figure~\ref{fig:noguess}.
It can be observed that the effect strongly depends on the initial solution given to the NLP solver. Therefore, heuristic is employed to remedy this problem, as explained in this section.

The idea of the heuristic is to find a set of key points, each of which is likely to yield a high reward.
Additionally, key points should approximate the global structure of the reward function.
A good initial solution can then be obtained by finding a path that passes through all the key points.
This allows the global structure to be exploited, rather than just optimizing locally as would be the case if the MPC were simply run without this step.

The presented algorithm is based on Gaussian mixture models (GMMs).
They can approximate any probability density function \cite{Goodfellow_2016} and have also been used before to prioritize search subregions, e.g., based on a probability map \cite{HierarchicalHeuristic,MoshagenEtAl:2025}.
Note that it is not necessary to consider the reward in \eqref{eq:integral} as a probability density function. However, we make the additional assumptions that the reward is always positive and that the total collectable reward is finite:
\begin{subequations}
\begin{equation}\label{eq:assump1}
    r(x) \geq 0 \quad \forall x \in \R^n\;,
\end{equation}
\begin{equation}\label{eq:assump2}
    \int r(x) \mathrm{d}x < \infty\;.
\end{equation}
\end{subequations}
Note that under assumptions (\ref{eq:assump1}) and (\ref{eq:assump2}), any reward function can be considered a probability density function (up to a scaling factor).

A Gaussian mixture with $n$ components is a function given by
\begin{equation}
p(x) = \sum_{k=1}^{n} w_{k} \mathcal{N}(x; \mu_{k}, \Sigma_{k})\;,
\end{equation}
where \(\mathcal{N}(x; \mu_{k}, \Sigma_{k})\) is a multivariate normal distribution and \(w_{k}\) are the weights of the components,] and \(n\) denotes the number of mixture components, i.e., the number of key points.

\begin{figure}
  \includegraphics[width=0.48\linewidth]{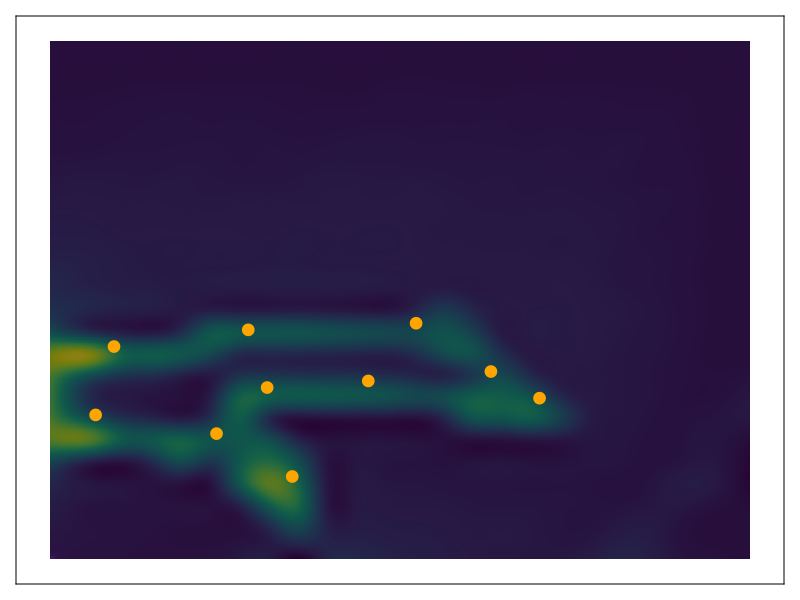}
  \includegraphics[width=0.48\linewidth]{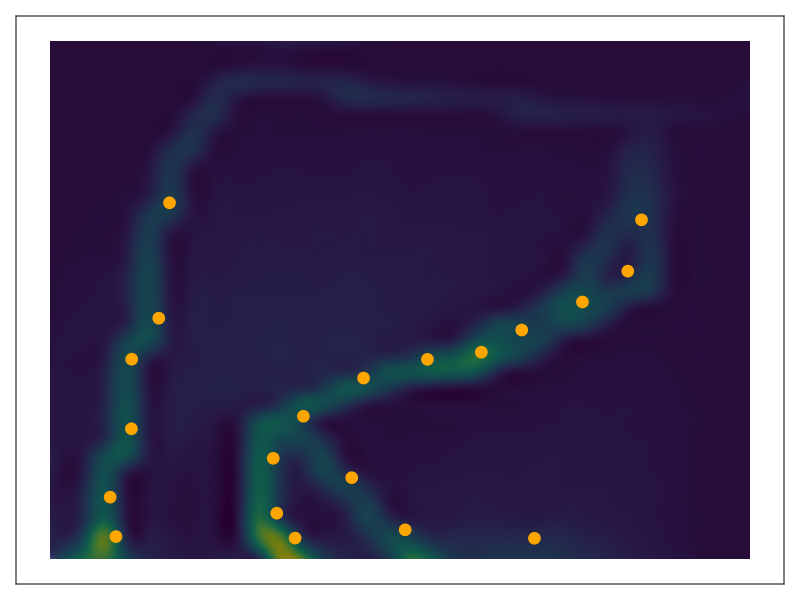}
  \caption{Key points found using Gaussian mixtures with $n=20$ ($m=10$) and $n=40$ ($m=20$) components respectively.}
  \label{fig:key-points}
\end{figure}

The parameters $w_{k},\mu_{k},\Sigma_{k}$ are optimized using the Expectation Maximization (EM) algorithm~\cite{ExpectationMaximization}.
Hence, by limiting the number of mixture components, the centers of the Gaussians are expected to be at positions yielding a high reward.
It can lead to better results if only the best (in terms of mean value) \(m < n\) components are used, for instance $m = n/2$.
In this case, we choose $m$ to be the number of key points. See Figure~\ref{fig:key-points} for results using this approach.

Given the set of key points, we now want to find the shortest path that passes all of them.
However, this problem is hard in itself, since even finding a shortest path through all key points without considering rewards is an instance of the TSP.
Although there are variants of TSP that also take into account the potential rewards for each city, the regular variant is used in the following.

\begin{figure}
  \includegraphics[width=\linewidth]{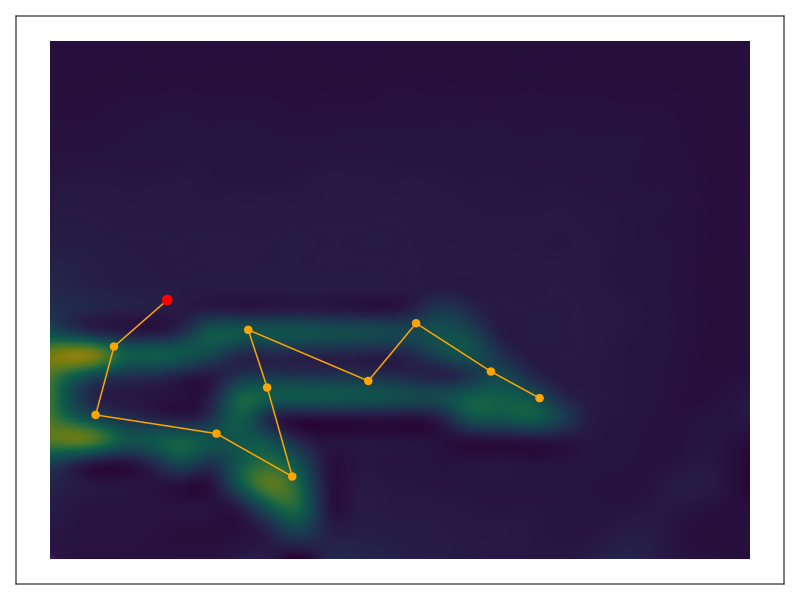}
  \caption{A travelling salesman tour through all key points (orange), starting from the initial position $x_{0}$ (red). The tour also ends in $x_{0}$, which is ommitted here.}
  \label{fig:tour}
\end{figure}

Typically, the TSP aims for a route returning back to the origin, which is not the desired outcome in this case.
To circumvent this issue, one can set the cost of travelling from any location to the current location to zero. Assuming $p_0$ is the current location, the cost function for two points $p_i$ and $p_j$ is given as
$$d(p_i,p_j) = \begin{cases}0 & \text{ if } j = 0\\\|p_{i}-p_{j}\|_{2} & \text{ otherwise}\end{cases}\;.$$
This forces the algorithm to start at the current position and returns the desired result.
The implementation uses a Dantzig–Fulkerson–Johnson formulation and is based on~\cite{Dantzig_1954,Pferschy_2016}. See Figure~\ref{fig:tour} for results.

\begin{figure}
\includegraphics[width=\linewidth]{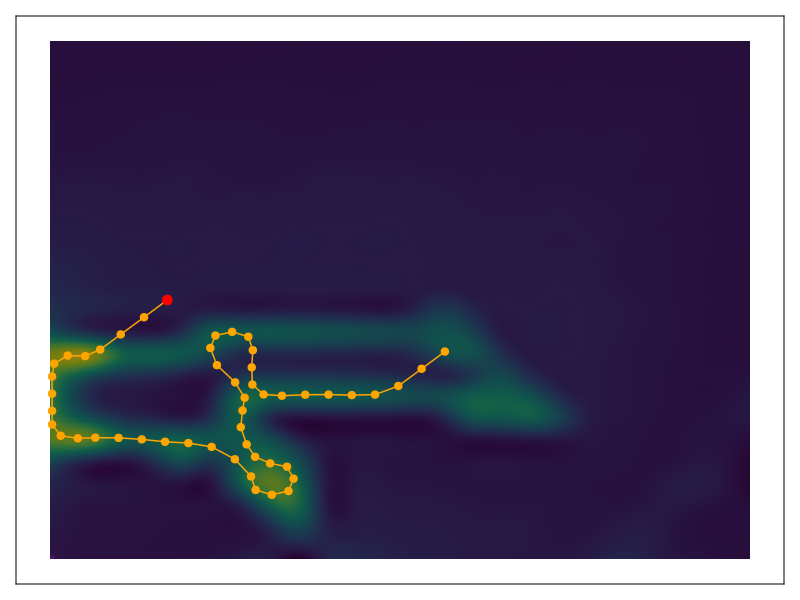}
\caption{The resulting open-loop trajectory of the MPC controller based on the initial solution obtained from the tour in Figure~\ref{fig:tour}.}
\label{fig:mpc}
\end{figure}

To feed the travelling salesman tour from the as an initial solution to the MPC problem \eqref{prob}, the tour is discretized using the maximum velocity as the step size. Figure~\ref{fig:mpc} shows the result of running the MPC as the closed-loop controller, given the tour from Figure~\ref{fig:tour} as the initial solution.

\section{Results}

The methods described in Section~\ref{methods} have been implemented in Julia, using the JuMP toolkit~\cite{JuMP} for defining optimization problems.
As the underlying solver, HiGHS~\cite{Highs} has been used for solving mixed-integer problems (TSP), and Ipopt~\cite{Ipopt} for non-linear continuous optimization problems (MPC).
The parameters of the objective function (\ref{prob:obj}) have been selected as \(c_{1} = 1, c_{2} = 1000, c_{3} = 100\).
All experiments have been performed on a machine with an Intel i5-1350P processor.

\begin{figure}
  \begin{subfigure}{\linewidth}
    \centering
    \includegraphics[width=0.48\textwidth]{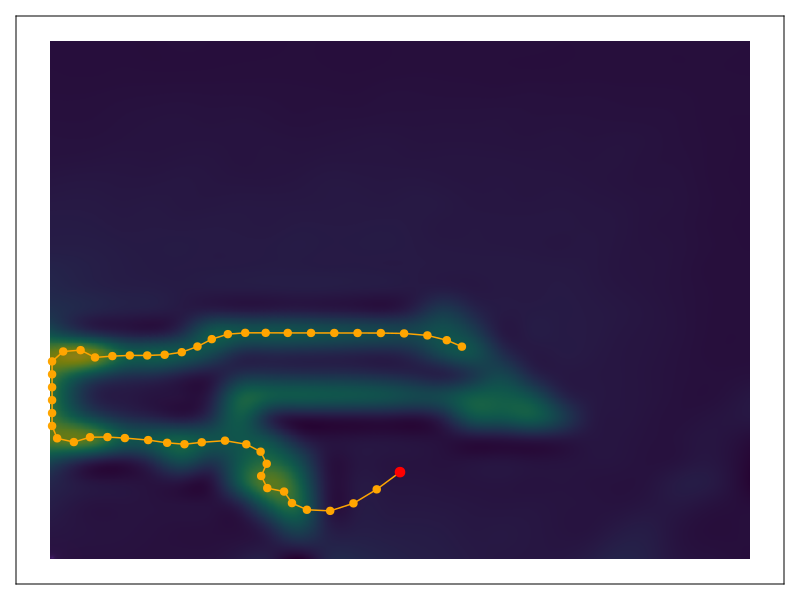}
    \includegraphics[width=0.48\textwidth]{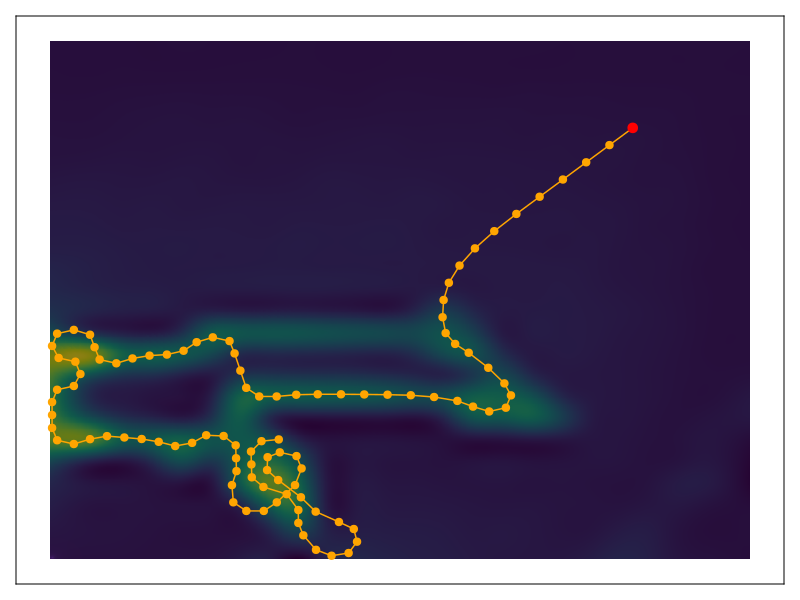}
    \caption{Scenario 1: $600 \times 600$ map} with horizon length $N=50$ (left) and $N=100$ (right)
    \label{fig:results:1}
  \end{subfigure}
  \begin{subfigure}{\linewidth}
    \centering
    \includegraphics[width=0.48\textwidth]{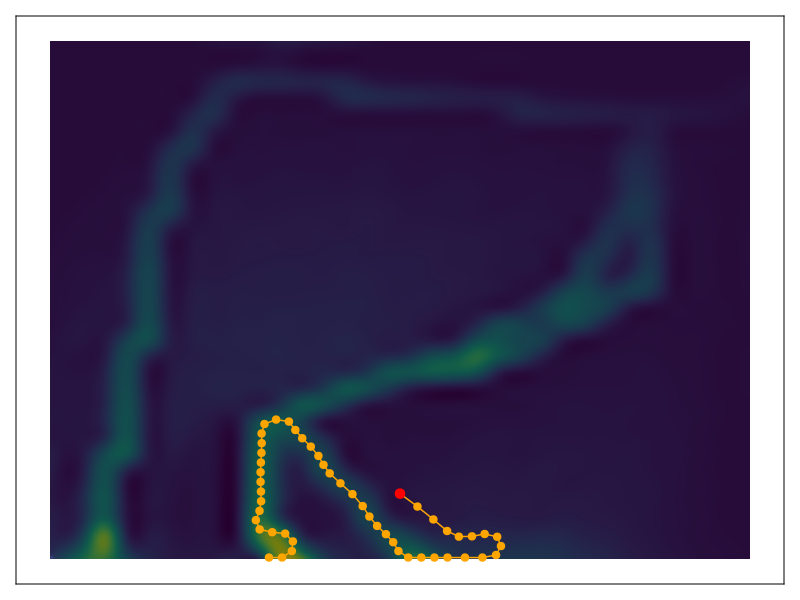}
    \includegraphics[width=0.48\textwidth]{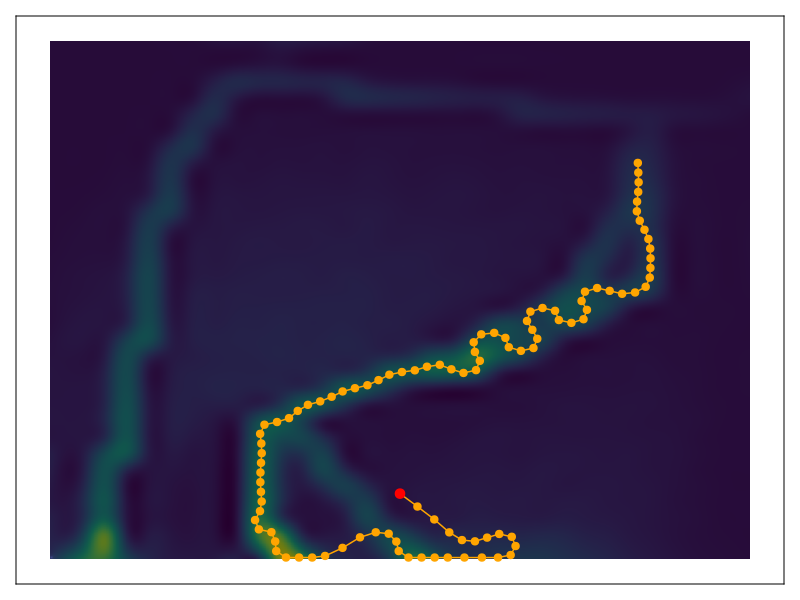}
    \caption{Scenario 2a: $800 \times 800$ map} with horizon length $N=50$ (left) and $N=100$ (right)
    \label{fig:results:2}
  \end{subfigure}
  \begin{subfigure}{\linewidth}
    \centering
    \includegraphics[width=0.48\textwidth]{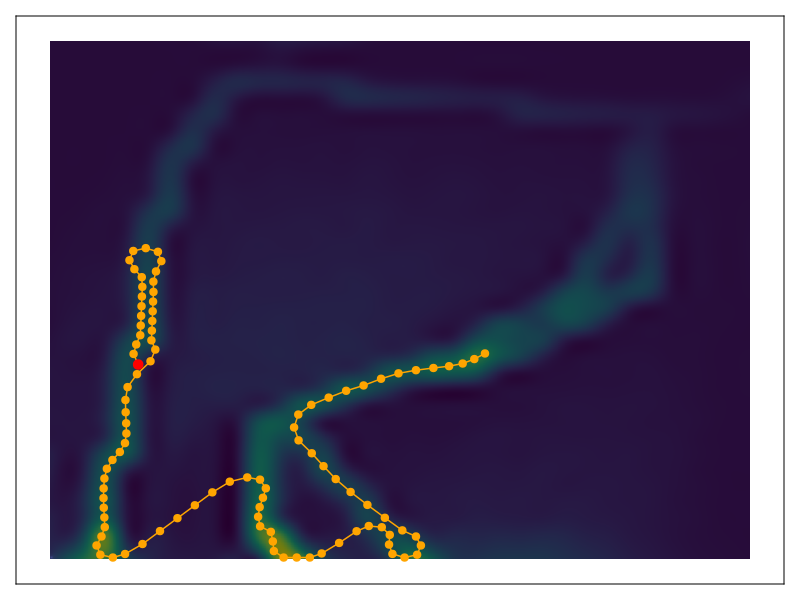}
    \includegraphics[width=0.48\textwidth]{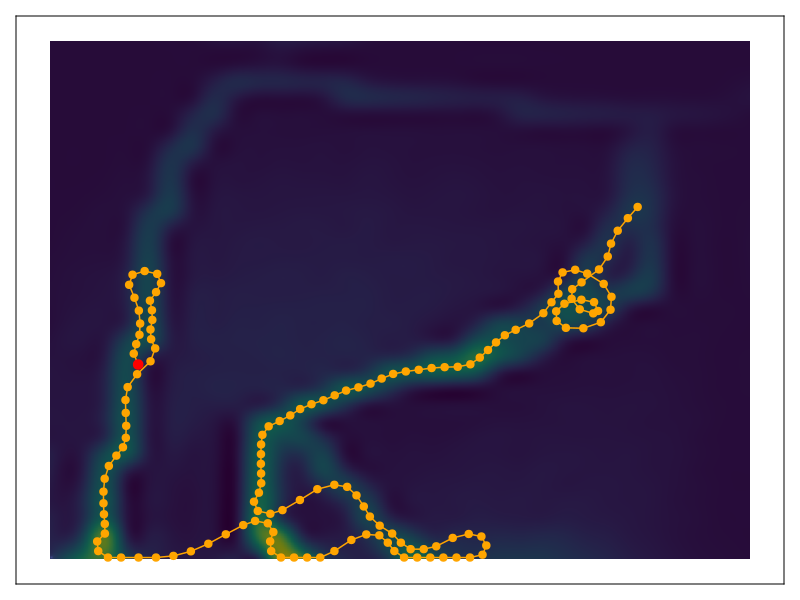}
    \caption{Scenario 2b: $800 \times 800$ map} with horizon length $N=100$ (left) and $N=150$ (right)
    \label{fig:results:2b}
  \end{subfigure}
  \begin{subfigure}{\linewidth}
    \centering
    \includegraphics[width=0.48\textwidth]{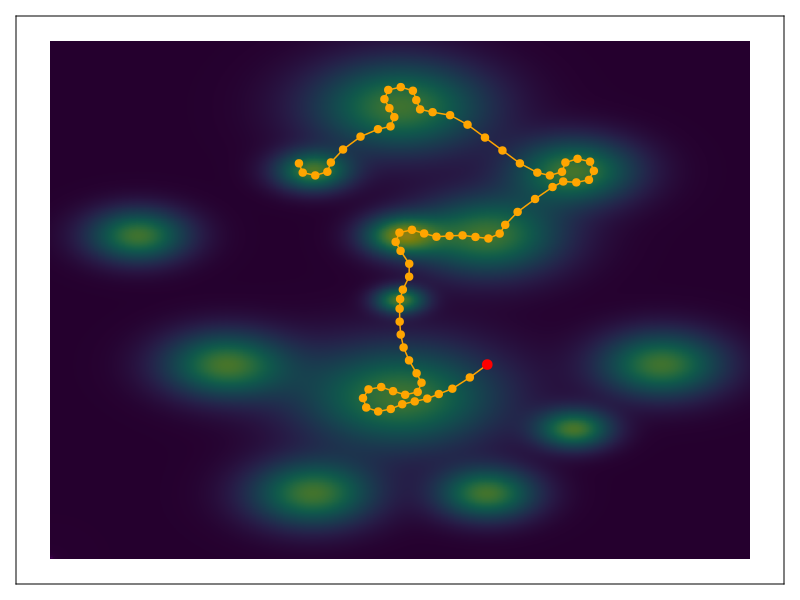}
    \includegraphics[width=0.48\textwidth]{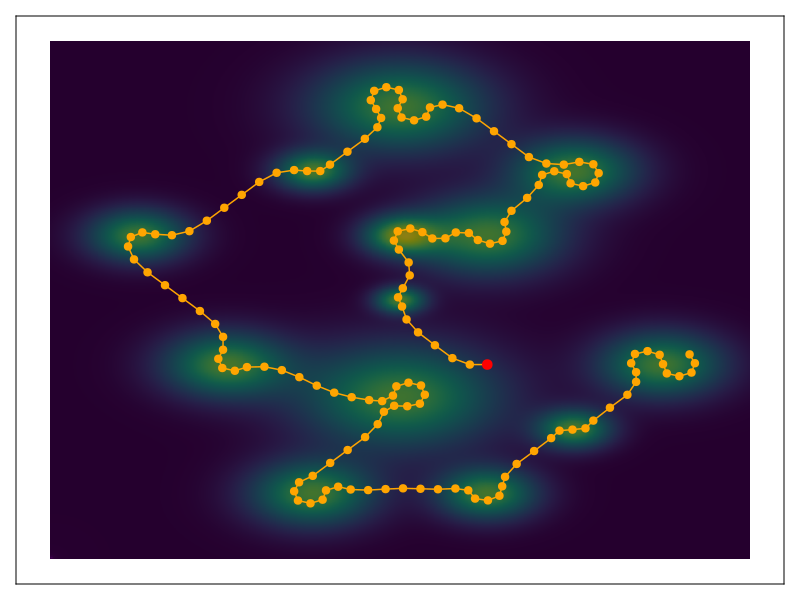}
    \caption{Scenario 3: $800 \times 800$ map with horizon length $N=75$ (left) and $N=150$ (right)}
    \label{fig:results:3}
  \end{subfigure}
  \caption{Results for three different scenarios, with increased horizon length on the right side. The red marker denotes the initial position of the UAV.}
  \label{fig:results}
\end{figure}

The implementation has been tested for scenarios in which a path is calculated for a UAV moving in a probability map, with the objective being to move through areas with the highest probability. Three different scenarios with different horizon lengths have been considered.
The first two scenarios in Figure~\ref{fig:results:1} and Figure~\ref{fig:results:2} are based on a given discrete probability map with sizes $600 \times 600$ meters and $800 \times 800$ meters, respectively.
Both maps are defined on a regular grid with a cell size of $25$ meters.
The initial position can be chosen freely.
To obtain a continuous reward function from the discrete probability map, a B-spline interpolation is used.
The third scenario in Figure~\ref{fig:results:3} is defined by combining multiple Gaussian distributions.

The runtime performance for all three scenarios is shown in Table~\ref{tab:table:runtime}.
As can be seen in the table, most of the time is spent on solving the nonlinear optimization problem.
Nevertheless, the quality of the solution compared to running the MPC without an initial guess could be improved in all secenarios.
This can be observed in Figure~\ref{fig:noguess}, which shows the solution for Scenario 1 but without an initial guess. 
The total reward value is lower, and the controller tends to stay in a local region instead of going to regions with high rewards.

\begin{table}[h]
\begin{tabular}{l|ll|llll}
 & $n$ & $N$ & GMM & TSP & MPC & Total \\
\hline
Scenario 1 & 20 & 50 & 0.18s & 0.004s & 6.42s & 6.6s \\
Scenario 1 & 20 & 100 & 0.23s & 0.006s & 28.39s & 28.63s \\
Scenario 2a & 40 & 50 & 0.41s & 0.072s & 3.31s & 3.79s \\
Scenario 2a & 40 & 100 & 0.51s & 0.27s & 38.15s & 38.93s \\
Scenario 2b & 40 & 100 & 0.4s & 0.03s & 16.43s & 16.86s \\
Scenario 2b & 40 & 150 & 0.4s & 0.049s & 53.9s & 54.35s \\
Scenario 3 & 40 & 75 & 0.37s & 0.007s & 4.1s & 4.48s \\
Scenario 3 & 40 & 150 & 0.62s & 0.012s & 41.89s & 41.52s \\
\end{tabular}
\caption{Runtime performance of the scenarios in Figure~\ref{fig:results}, where $n$ is the number of mixture components, and $N$ is the horizon length of the MPC}
\label{tab:table:runtime}
\end{table}

\section{Conclusion}

The paper presents a new approach of using MPC for the WCPP problem, where a spatially distributed reward is collected by an agent with continuous dynamics. To this end, the new concept of CCs is introduced.

The performance of the solution is improved by providing an initial guess based on the solution of a TSP, which finds the optimal path going through a set of key points obtained by approximating the reward function using GMMs.

Specifically, we consider path planning for SAR missions, where the reward is given as the probability of finding a missing person in a given area or map. Correspondingly, the proposed approach has been evaluated on a number of specific problem instances, demonstrating its practical and computational viability.

Ross et al. showed that TSP problems can also be solved using optimal control formulations \cite{Ross_2020a}, so it may be possible to merge the steps from Sections \ref{impl2} into a single formulation in a future work. Furthermore, it could be interesting to try other MPC formulations and/or CCs and evaluate them in the context of the WCPP. 

\bibliographystyle{ieeetr}
\bibliography{references}

\end{document}